\documentclass[a4paper,11pt]{article}
\usepackage{pos}

\newcommand{\apj}{Astrophysical Journal}
\newcommand{\apjl}{Astrophysical Journal Letters}
\newcommand{\prd}{Physical Review D}
\newcommand{\apjs}{Astrophysical Journal Supplement}
\newcommand{\aap}{Astronomy and Astrophysics}
\newcommand{\aj}{Astronomical Journal}

\newcommand{\mnras}{Monthly Notices of the RAS}

\manuallySeparateAuthors

\title{Gamma-ray emission from young radio galaxies and quasars: the flaring episode of the peculiar galaxy PKS\,B1413+135}
 \ShortTitle{Gamma-rays from young radio galaxies and quasars}

\author*[a,b,c]{Giacomo Principe}
\author[d]{, Leonardo Di Venere}
\author[c]{, Giulia Migliori}
\author[c]{, Monica Orienti}
\author[c]{, Filippo D'Ammando}
\author{, on behalf of the Fermi Large Area Telescope Collaboration.}

\affiliation[a]{Universit\'a di Trieste, Dipartimento di Fisica,\\  I-34127 Trieste, Italy}

\affiliation[b]{Istituto Nazionale di Fisica Nucleare, Sezione di Trieste,\\ I-34127 Trieste, Italy}

\affiliation[c]{Istituto Nazionale di Astrofisica - Istituto di Radioastronomia,\\ I-40129 Bologna, Italy}

\affiliation[d]{Istituto Nazionale di Fisica Nucleare, Sezione di Bari,\\ I-70126 Bari, Italy}

\emailAdd{giacomo.principe@ts.infn.it}

\abstract{According to radiative models, radio galaxies are predicted to produce gamma rays from the earliest stages of their evolution onwards.The study of the high-energy emission from young radio sources is crucial for providing information on the most energetic processes associated with these sources, the actual region responsible for this emission, as well as the structure of the newly born radio jets. 
Despite systematic searches for young radio sources at gamma-ray energies, only a handful of detections have been reported so far. Taking advantage of more than 11 years of \textit{Fermi}-LAT data, we investigate the gamma-ray emission of 162 young radio sources (103 galaxies and 59 quasars), the largest sample of young radio sources used so far for a gamma-ray study. We analyse the \textit{Fermi}-LAT data of each individual source separately to search for a significant detection. In addition, we perform the first stacking analysis of this class of sources in order to investigate the gamma-ray emission of the young radio sources that are undetected at high energies. 
We report the detection of significant gamma-ray emission from 11 young radio sources, including the discovery of significant gamma-ray emission from the compact radio galaxy PKS 1007+142. 
Although the stacking analysis of below-threshold young radio sources does not result in a significant detection, it provides stringent upper limits to constrain the gamma-ray emission from these objects.
In this talk we present the results of our study and we discuss their implications for the predictions of gamma-ray emission from this class of sources.}

\FullConference{37$^{\rm{th}}$ International Cosmic Ray Conference (ICRC 2021)\\
		July 12th -- 23rd, 2021\\
		Online -- Berlin, Germany}

\begin{document}
\maketitle

\section{Introduction}
One of the greatest challenges faced by modern astrophysics is understanding the origin of the gamma-ray emission in radio galaxies and quasars.
While the extragalactic gamma-ray sky is dominated by blazars only few radio galaxies have been detected so far \citep[4LAC,][]{2020ApJ...892..105A}. With their misaligned jets, they offer a unique tool to probe some of the non-thermal processes at work in unbeamed regions in AGN.

According to the evolutionary scenario, the size of a radio galaxy is strictly related to its age \citep{1995A&A...302..317F}. Therefore extragalactic compact radio objects (projected linear size LS<20 kpc), 
are expected to be the progenitors of extended radio galaxies \citep{1996ApJ...460..612R}.
A support to the young nature of these object is given by the determination of kinematic and radiative ages in some of the most compact sources ($t \sim 10^2 - 10^5$ years) \citep{1982A&A...106...21P}. Compact radio objects reside either in galaxies or quasars. While for the latest the gamma-ray emission is favored by their smaller jet-inclination angle and beaming effects, the origin of gamma-ray emission in galaxies is still a matter of debate.

\citet{2008ApJ...680..911S} predicted that young radio galaxies would constitute a large class of sources detectable by \textit{Fermi}-LAT. They are expected to produce isotropic $\gamma$-ray emission through IC scattering of the UV photons by the electrons in the compact radio lobes. 
The search for high-energy emission from young radio galaxies and quasars is crucial for investigating the energetic processes in the central region of the host galaxy, as well as the origin and the structure of the newly born radio jets.
However, systematic searches of $\gamma$-rays from young radio sources have so far been unsuccessful \citep{2016AN....337...59D}.
Dedicated studies reported a handful of detections: three young radio galaxies (NGC\,6328 \citep{2016ApJ...821L..31M}, NGC\,3894 \citep{2020A&A...635A.185P} and TXS\,0128+554 \citep{2020ApJ...899..141L}),
and five compact steep spectrum (CSS) sources (3C\,138, 3C\,216, 3C\,286, 3C\,380, and 3C\,309.1) all associated with quasars \citep[4FGL,][]{2020ApJS..247...33A}.

Taking advantage of the increased sensitivity provided by more than eleven years of LAT data, we investigate the gamma-ray properties of a sample of 162 young radio sources (103 galaxies and 59 quasars). In particular in the second part of this proceeding we investigate the high-energy emission of PKS\,B1413+135 and discuss the nature of this peculiar galaxy.



\section{Sample of young radio sources and analysis description}

\label{sec:sample}
In order to choose young radio sources, we base our sample on the following resources which contain radio galaxies and quasars with projected linear size below 50 kpc: \citet{2009A&A...498..641D,2014MNRAS.438..463O, 2020MNRAS.491...92L, 2020ApJ...892..116W}. 
We added to our sample also the sources NGC\,3894, TXS\,0128+554, and 3C\,380, since already detected at high energy and investigated in \citet{2020A&A...635A.185P}, \citet{2020ApJ...899..141L} and \citet{2020ApJ...899....2Z}, respectively. Our final sample consists of 162 young radio sources (103 galaxies and 59 quasars) with known position, redshift, linear size, radio luminosity and peak frequency.


\subsection{Analysis Description}
\label{sec:analysis_single_source}
We performed a dedicated analysis of each individual source of our sample using more than 11 years of \textit{Fermi}-LAT data between August 5, 2008, and November 1, 2019. We selected LAT data from the P8R3 Source class events \citep{2018arXiv181011394B}, and P8R3\_Source\_V2 instrument response functions (IRFs), in the energy range between 100\,MeV and 1\,TeV, in a region of interest (ROI) of 20$^{\circ}$ radius centered on the source position. 
The lower limit for the energy threshold is driven by the large uncertainties in the arrival directions of the photons below 100 MeV, which may be confused with the Galactic diffuse component. See \citet{2018A&A...618A..22P, 2019RLSFN.tmp....7P} for a different analysis technique to solve this and other issues at low energies with \textit{Fermi}-LAT.

The analysis procedure applied in this work is mainly based on two steps. First, we investigate the $\gamma$-ray data of each individual source with a standard likelihood analysis \citep[see e.g.][]{2020A&A...635A.185P,2021ApJ...911L..11E}.
The likelihood analysis consists of model optimization, source localization, spectrum and variability study. It was performed with \texttt{Fermipy}\footnote{http://fermipy.readthedocs.io/en/latest/} \citep{2017ICRC...35..824W}. The counts maps were created with a pixel size of $0.1^{\circ}$. We selected $\gamma$-rays with zenith angle smaller than 105$^{\circ}$ (95$^{\circ}$, 85$^{\circ}$) and all event types\footnote{A measure of the quality of the direction reconstruction assigns events to four quartiles, the PSF event types: 0, 1, 2, 3, where PSF0 has the largest point spread function and PSF3 has the best.} (excluding PSF0, excluding PSF1) in order to limit the contamination from the Earth’s limb \citep{2009PhRvD..80l2004A} at energies above 1 GeV (between 300 MeV and 1 GeV, below 300 MeV, respectively). The model used to describe the ROI includes all point-like and extended LAT sources, located at a distance $<25^{\circ}$ from each target position, listed in 4FGL catalog \citep{2020ApJS..247...33A}, as well as the Galactic diffuse and isotropic emission \footnote{https://fermi.gsfc.nasa.gov/ssc/data/access/lat/BackgroundModels.html}.
For the analysis we first optimized the model for the ROI, then we searched for the possible presence of new sources and finally we re-localized the source. During the model optimization, we left free to vary the diffuse background and all the spectral parameters of the sources within 5$^{\circ}$ of our targets, and only the normalization for those at a distance between 5$^{\circ}$ and 10$^{\circ}$. To perform a study of the $\gamma$-ray emission variability of each source we divided the \textit{Fermi}-LAT data into time intervals of one year and left only the normalization free to vary. 

In addition to the study of each individual source, we performed a stacking analysis of the sources which were not significantly detected, in order to investigate the general properties of the
population of young radio galaxies and quasars.


\section{\textit{Fermi}-LAT results}
\label{sec:results}

\begin{figure}
\centering
\includegraphics[width=14cm]{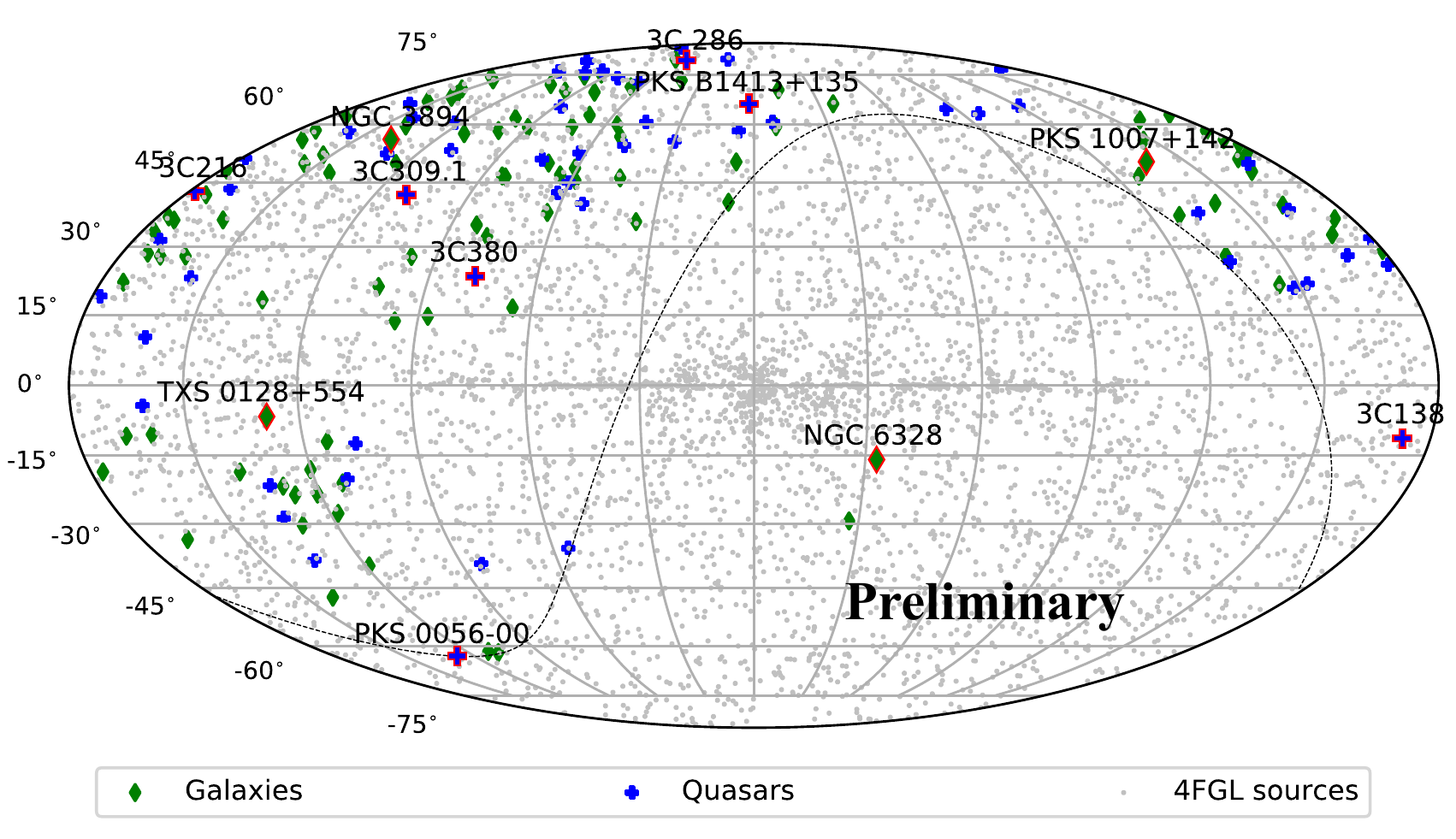}
\caption{\small \label{fig:sky_map}
Sky map, in Galactic coordinates and Mollweide projection, showing the young radio sources in our sample. The detected sources are labelled in the plot. All the 4FGL sources \citep{2020ApJS..247...33A} are also plotted, with grey points, for comparison.}
\end{figure}

In our study we detect significant $\gamma$-ray emission (TS $>$ 25) at the positions of 11 young radio sources (see Fig. \ref{fig:sky_map}), four galaxies: NGC\,6328 (z=0.014, LS=2pc), NGC\,3894 (z=0.0108, LS=10 pc), TXS\,0128+554 (z=0.0356, LS=12 pc) and PKS\,1007+142 (z=0.213, LS= 3.3 kpc);  and seven quasars: 3C\,138 (z=0.759, LS= 5.9 kpc), 3C\,216 (z=0.6702, LS=56kpc), 3C\,286 (z=0.85,  LS=25 kpc), 3C\,309.1 (z=0.905, LS= 17 kpc), 3C\,380 (z=0.692,LS= 11 kpc), PKS\,0056-00 (z=0.247, LS= 0.03) and PKS\,B1413+135 (LS=0.247, 0.03 kpc)

Nine out of the 11 detected sources were present in previous \textit{Fermi}-LAT catalogs, while PKS\,0056-00 has been recently reported in the latest release of LAT sources 4FGL-DR2\footnote{https://fermi.gsfc.nasa.gov/ssc/data/access/
lat/10yr\_catalog/} \citep{2020ApJS..247...33A}.
In addition to the sources already included in the 4FGL-DR2, we report here the discovery of $\gamma$-ray emission from the young radio galaxy PKS\,1007+142 ($z$ = 0.213). 

The stacking analysis on the LAT data of the undetected sources did not result in the detection of significant emission. The upper limits obtained with this procedure are, however, about one order of magnitude below than those derived from the individual sources. This allow us a comparison with the model proposed by \citet{2008ApJ...680..911S}, predicting isotropic gamma-ray emission from the compact lobes of young radio galaxies, excluding jet powers $(\gtrsim$10$^{42}$--10$^{43}$ erg s$^{-1}$) coupled with UV luminosities $>$ 10$^{45}$ erg s$^{-1}$. More information on this project can be found in \citep{2021MNRAS.507.4564P}.  

In the following,we investigate the high-energy emission of PKS\,B1413+135, and discuss the nature of this peculiar galaxy.

\section{The peculiar galaxy PKS\,B1413+135}
It is an unusual object whose classification has been longly debated. It is a very compact (LS$\sim$30 pc) and young ($\tau \sim$100 years) radio source with a core-dominated structure, situated in a spiral galaxy at redshift $z$ = 0.247 \citep{1992ApJ...400L..13C,2017ApJ...845...90V}. Rapid variability at radio and infrared wavelengths, strong and rapid variable infrared polarization and a flat–spectrum radio continuum, first lead to a classification of PKS\,1413+135 as a BL Lac object \citep{1992ApJ...400L..17S}. Subsequently, radio observations using VLBI techniques has questioned this classification. The radio source presents: (1) a compact radio core, (2) a jet–like structure on a parsec scale, and (3) a counter–jet \citep{1996AJ....111.1839P,2005ApJ...622..136G}. The presence of a counter–jet disagrees with the interpretation of PKS\,B1413+135 as a BL Lac object, since these are believed to represent radio sources with the jet pointing almost exactly along the line–of–sight, with the flux boosted by relativistic beaming. The counter–jet would not be visible in this scenario.


\subsection{\textit{Fermi}-LAT results on PKS 1413+135}
\label{fermi_results}
\begin{figure}[h]
\centering
\includegraphics[width=0.7\columnwidth]{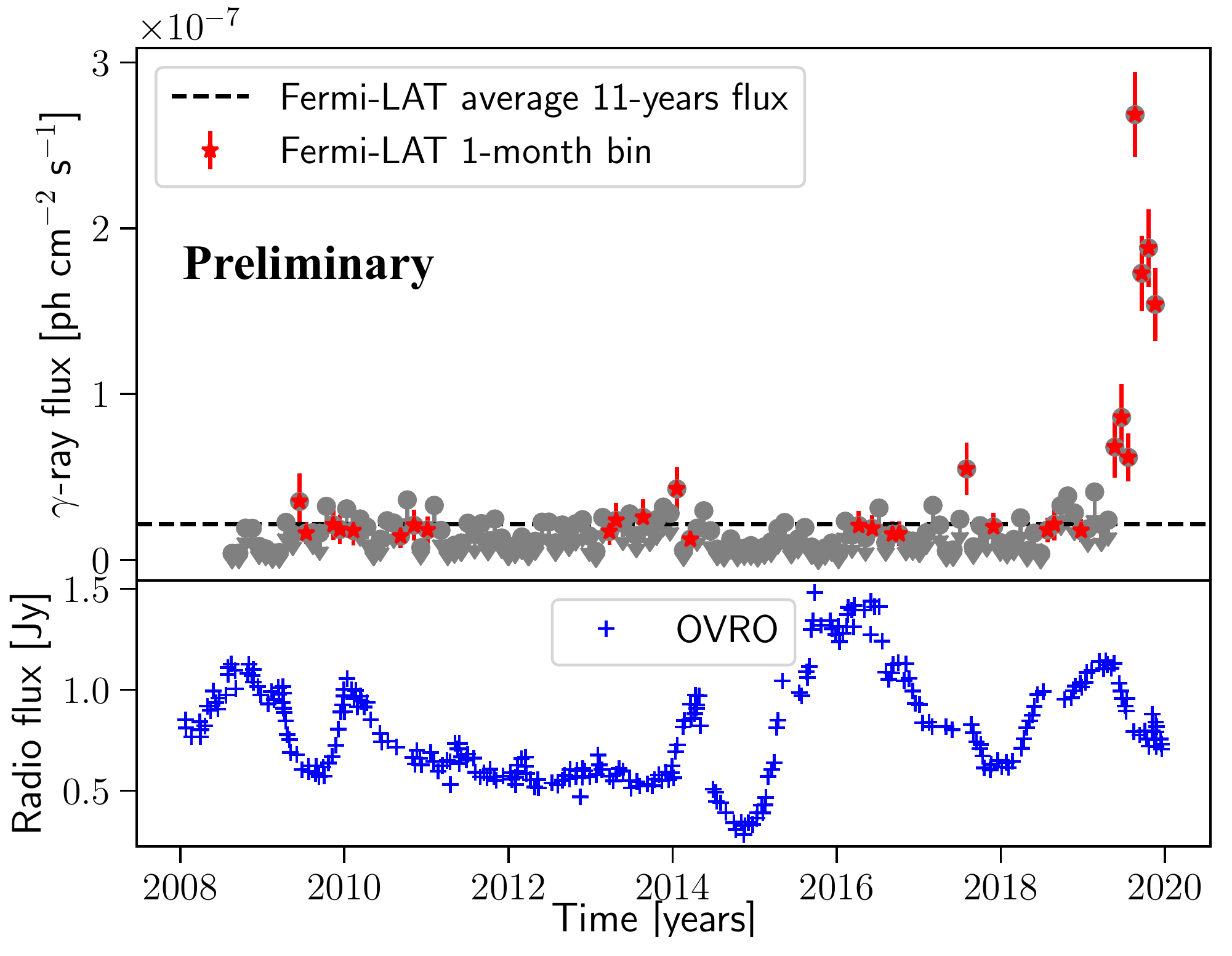}
\caption{\small \label{PKSB1413_flare} Top: \textit{Fermi}-LAT one-month binned light curve of PKS\,B1413+135. The dashed line represents the averaged flux for the entire period $Flux_{11\,years}=1.47\times 10^{-8}$ ph cm$^{-2}$ s$^{-1}$. The 95\% upper limit are reported in each time interval with TS $<10$. Bottom: radio lightcurve from the OVRO 40 m Telescope 15 GHz monitoring program (https: //www.astro.caltech.edu/ovroblazars/) \citep{2021ApJ...907...61R}.}
\end{figure}

PKS\,B1413$+$135 is significantly detected (TS = 2486) in our analysis, showing extremely bright $\gamma$-ray emission, with an average flux $F = (1.47 \pm 0.1) \times 10^{-8}$ ph cm$^{-2}$ s$^{-1}$  and a power-law index for PKS\,B1413+135 is $\Gamma = 2.10 \pm0.03$. To perform a study of the $\gamma$-ray emission variability of PKS\,B1413+135, we divided the \textit{Fermi}-LAT data into time intervals of one month. For the light-curve analysis we fixed the photon index to the value obtained for 11 years of data, and left only the normalization free to vary. Figure \ref{PKSB1413_flare} shows the \textit{Fermi}-LAT light curve for one-month time bins, compared to the 15 GHz radio lightcurve taken by the OVRO 40 m Telescope \citep{2021ApJ...907...61R}.



Its light curve reveals a strong $\gamma$-ray flare in the last year (2019) considered in this analysis. The flux reached the peak on August 29, 2019, when the source was detected with a significance of TS = 67. We measured a daily flux of $F_{E > 100 \,\textrm{MeV}} = (5.4 \pm 1.9) \times 10^{-7}$ ph cm$^{-2}$ s$^{-1}$ and a significant hardening of the spectrum: $\Gamma = 2.0 \pm 0.2$ ($\Gamma_{\rm\,4FGL} = 2.41 \pm 0.07$), in agreement with the preliminary results reported by \citet{2019ATel13049....1A}. 

\subsection{The flaring episode of PKS B1413+135}
\begin{figure}[h]
\centering
\includegraphics[width=0.75\columnwidth]{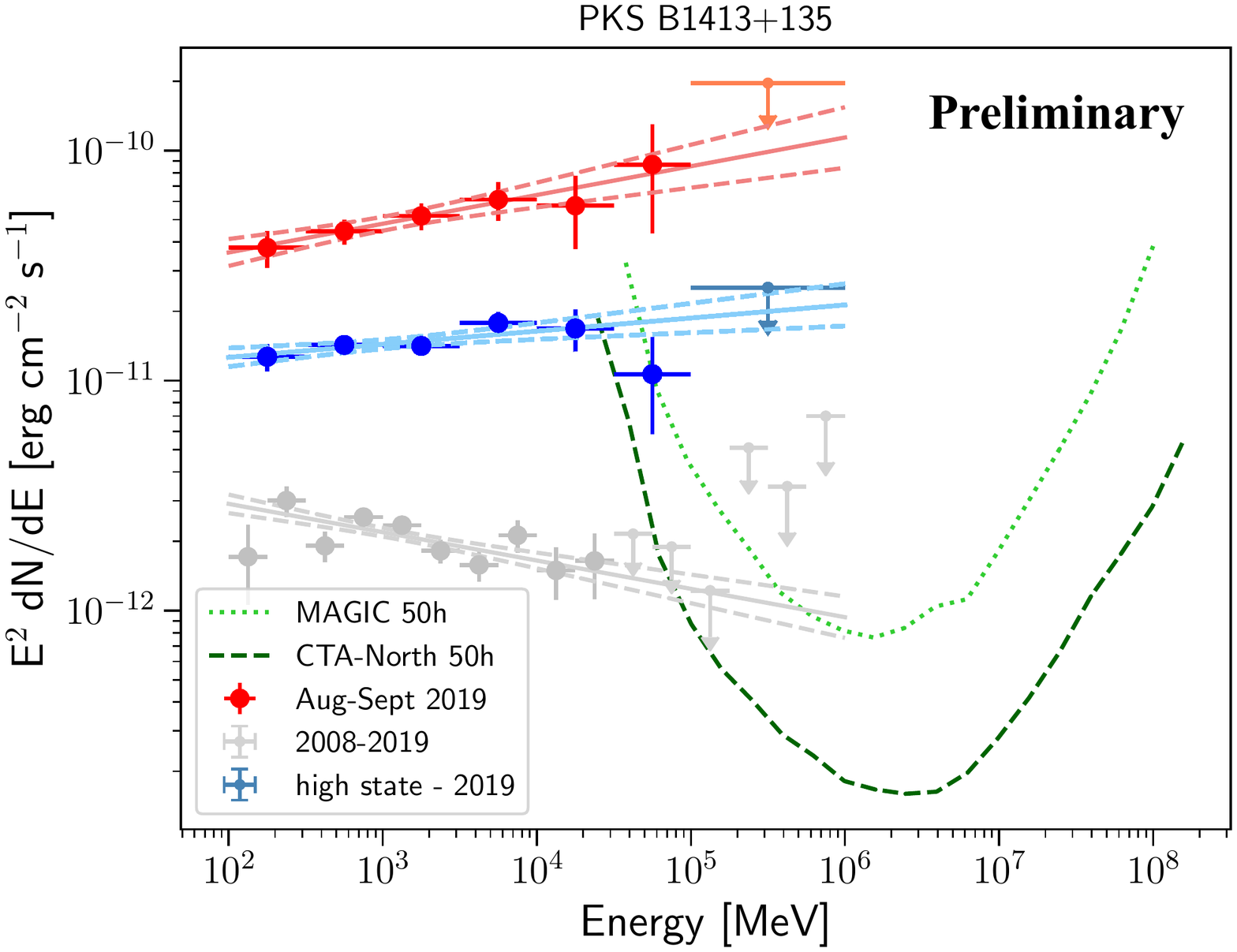}
\caption{\small \label{PKSB1413_sed_flare} \textit{Fermi}-LAT spectrum of PKS\,B1413+135 during the high state emission (2019) and the one-month interval centered on the peak of flaring emission (August 29, 2019). For a comparison the SED obtained with the 12 years is plotted in light-grey. MAGIC 50 hours sensitivity \citep{2016APh....72...76A} and preliminary CTA 50 hours sensitivity for the Alpha configuration \citep[see EAS 2021 talk of Juan Cortina and ][]{2019scta.book.....C} are also plotted for a comparison.}
\end{figure}

In order to investigate the $\gamma$-ray emission during its bright flaring episode, we performed a dedicated analysis of the entire high state period (the ten-months time interval between January 1, 2019 and November 5, 2019), as well as on the one-month period centered on the peak of the flaring episode (August 15, September 15, 2019). Figure \ref{PKSB1413_sed_flare} shows the spectral energy distribution (SED) for 11 years of \textit{Fermi}-LAT data, as well as for the high-state period and the one-month interval centered on the peak of the flaring episode. 

During 2019 the source is significantly detected (TS=1778) and it presents a bright flux of $F_{2019}=(8.33\pm 0.57) \times 10^{-8}$ ph cm$^{-2}$ s$^{-1}$, about six times larger than the averaged emission on the 12 years. In particular, the source shows a significant spectral hardening during its high state, with a power index $\Gamma_{2019}= 1.94\pm 0.03$. 
Focusing on the one month time interval centered on the peak of the flaring emission (August 29, 2019) the source presents an even brighter flux of $F_{Aug-Sept,2019}= (2.57\pm 0.25) \times 10^{-7}$ ph cm$^{-2}$ s$^{-1}$, and harder spectral index $\Gamma_{Aug-Sept,2019}= 1.86\pm0.05$.
Due to its bright emission and spectral hardening, the $\gamma$-ray emission of the flaring episode of PKS\,B1413+135 would be detectable also at very high energies (VHE, E$>50$ GeV) by the current generation of imaging air cherenkov telescopes (IACT, e.g. MAGIC) within 50 hours \citep{2016APh....72...76A}, while the next cherenkov telescope array\footnote{CTA instrument response functions
provided by the CTA Consortium and Observatory, see https://www.cta-
observatory.org/science/cta-performance/ (version prod3b-v2) for more details.} (CTA) will be able to detected a similar flaring emission in an even shorter time scale  \citep{2019scta.book.....C}.

\section{Conclusion}

The goal of our study was to investigate the $\gamma$-ray properties of young-radio sources. To this end we analysed 11.3 years of \textit{Fermi}-LAT data for a sample of 162 sources. We analysed the $\gamma$-ray data of each source individually to search for a significant detection. We report the detection of significant $\gamma$-ray emission from 11 young radio sources, including the discovery of significant $\gamma$-ray emission from the compact radio galaxy PKS 1007+142. In addition, we perform the first stacking analysis of this class of sources in order to investigate the $\gamma$-ray emission of the young radio sources that are undetected at high energies, without finding significant emission. More information on this project can be found in \citep{2021MNRAS.507.4564P}.  

PKS\,B1413$+$135 has long been considered an unusual object, with a BL-Lac-like AGN hosted in a spiral galaxy at redshift $z$ = 0.247 \citep{1992ApJ...400L..13C,2017ApJ...845...90V}. 
In our analysis we observed a $\gamma$-ray flare at the end of 2019 with a flux increase of a factor of 6 with respect to the average value, and up to a factor of 35 if we consider the daily peak flux. These kinds of flux increases have been observed during high activity states in many FSRQs and BL Lacs \citep{2010ApJ...722..520A}.
The detection of this flaring activity supports the idea that the $\gamma$-ray emission is beamed and produced by a relativistic jet at a relatively small viewing angle, similar to the case of the AGN PKS\,0521$-$36 \citep{2015MNRAS.450.3975D}. For the quasar PKS\,B1413$+$135 a recent study by \citet{2021ApJ...907...61R} argues that the association with the spiral host galaxy is just due to a chance alignment, and instead supports the hypothesis that PKS\,B1413+135 is a background blazar-like object lying in the redshift range  0.247 $<$ z $<$ 0.5.

In addition to a brightening of the emission we observed a significant spectral hardening, making the flaring episode of this gravitationally-lensed quasar PKS\,B1413+135 detectable also at VHE by the current IACT (e.g. MAGIC) within 50 hours, while CTA will be able to detected a similar flaring emission in an even shorter time. In the future, VHE observations of flaring activity of PKS\,B1413+135 by current IACT telescopes or by more sensitive instruments like CTA, are of fundamental interest as they allow us to probe the inner-jet environment and to study the possible link between VHE activity and jet formation in AGN.

\subsection*{ACKNOWLEDGMENTS}
The \textit{Fermi}-LAT Collaboration acknowledges support from NASA and DOE (United States), CEA/Irfu, IN2P3/CNRS, and CNES (France), ASI, INFN, and INAF (Italy), MEXT, KEK, and JAXA (Japan), and the K.A. Wallenberg Foundation, the Swedish Research Council, and the National Space Board (Sweden).

%

\end{document}